# Transactive Energy Auction with Hidden User Information in Microgrid

M.Nazif Faqiry, *IEEE Student Member*, Sanjoy Das

*Abstract—* This research proposes a novel auction mechanism for transactive energy exchange between buyers and sellers, modeled as agents in a microgrid. The mechanism is implemented by a separate microgrid controller (MC) agent, and requires big data flow with the other agents through an iterative bidding process. Although private user information remains hidden to the MC, a theoretical analysis shows that under the assumption of convexity of the agents' utilities, the proposed auction is still able to maximize the social welfare (SW), i.e. the aggregate utilities of the agents. In addition, it is shown that the mechanism exhibits the key desirable features of individual rationality and weak budget balance; guaranteeing that neither the payoff to any agent nor the net monetary revenue after termination, is negative. The proposed approach also incorporates a mechanism to redistribute the sellers' shares in a fair manner. As an example, a maximum entropy based fair redistribution scheme is addressed. The theoretical analysis reported here is accompanied by extensive set of simulations that illustrate the various aspects of the proposed mechanism.

## NOMENCLATURE

| | |
|---|---|
| $\mathcal{D}$ | Set of buyer agents |
| $\mathcal{S}$ | Set of seller agents |
| $N_b$ | Number of buyer agents, where $N_b = \|\mathcal{D}\|$ |
| $N_s$ | Number of seller agents, where $N_s = \|\mathcal{S}\|$ |
| $i$ | Index of a buyer, where $i \in \mathcal{D}$ |
| $j$ | Index of a seller, where $j \in \mathcal{S}$ |
| $u_i$ | Utility function of the $i^{th}$ buyer |
| $u_i'$ | Marginal utility of the $i^{th}$ buyer |
| $g_j$ | Generation capacity of the $j^{th}$ seller |
| $v_j$ | Utility function of the $j^{th}$ seller |
| $v_j'$ | Marginal utility of the $j^{th}$ seller |
| $d_i$ | Demand delivered to the $i^{th}$ buyer |
| $b_i$ | Buying price bid placed by the $i^{th}$ buyer |
| $c_i$ | Buying per unit price payed by the $i^{th}$ buyer |
| $a_j$ | Availability declared by $j^{th}$ seller |
| $s_j$ | Supply amount assigned to the $j^{th}$ seller |
| $c_j$ | Minimum per unit selling price by the $j^{th}$ seller |
| $p$ | The minimum per unit buying and maximum per unit selling price of energy |
| $\Theta$ | SW optimization problem (SWOP) objective function |
| $\mathcal{L}_\Theta$ | Lagrangian of the SWOP |
| $\lambda_i$ | Dual variable in $\mathcal{L}_\Theta$, corresponding to $pd_i \leq b_i$ |
| $\alpha_j$ | Dual variable in $\mathcal{L}_\Theta$, corresponding to $s_j < a_j$ |
| $\mu$ | Dual variable in $\mathcal{L}_\Theta$, corresponding to $\sum_{i \in \mathcal{D}} d_i = \sum_{j \in \mathcal{S}} s_j$ |
| $d_i^*$ | $d_i$ at equilibrium as the efficient solution of the SWOP |
| $s_j^*$ | $s_j$ at equilibrium as the efficient solution of the SWOP |
| $\lambda_i^*$ | $\lambda_i$ at equilibrium in the SWOP |
| $\alpha_j^*$ | $\alpha_j$ at equilibrium in the SWOP |
| $\mu^*$ | $\mu$ at equilibrium in the SWOP |
| $\Phi$ | MC optimization problem (MCOP) objective function |
| $\mathcal{L}_\Phi$ | Lagrangian of the MCOP |
| $\gamma_i$ | Dual variable in $\mathcal{L}_\Phi$, corresponding to $pd_i \leq b_i$ |
| $\beta_j$ | Dual variable in $\mathcal{L}_\Phi$, corresponding to $s_j < a_j$ |
| $\nu$ | Dual variable in $\mathcal{L}_\Phi$, corresponding to $\sum_{i \in \mathcal{D}} d_i = \sum_{j \in \mathcal{S}} s_j$ |
| $d_i^\dagger$ | $d_i$ at equilibrium as the solution of the MCOP |
| $s_j^\dagger$ | $s_j$ at equilibrium as the solution of the MCOP |
| $\gamma_i^\dagger$ | $\gamma_i$ at equilibrium in the MCOP |
| $\beta_j^\dagger$ | $\beta_j$ at equilibrium in the MCOP |
| $\nu^\dagger$ | $\nu$ at equilibrium in the MCOP |
| $\zeta$ | Dual variable of constraint $s_j \geq 0$ in seller's problem |
| $\pi_i$ | Buyers payoff from the auction |
| $\pi_j$ | Sellers payoff from the auction |
| $\pi_{MC}$ | Microgrid controllers' benefit |
| $F$ | Fairness term function |
| $\eta$ | Fairness term coefficient |
| $\mathcal{L}_{WF}$ | Lagrangian of the FROP |
| $s_j^r$ | The sellers' redistributed supply |
| $S$ | Sum of the redistributed supply of all sellers |
| $R$ | Total sellers' revenue |
| $c_j^r$ | The sellers' redistributed selling price |
| $\beta_j^r$ | Dual variable in $\mathcal{L}_{WF}$ for $s_j^r \leq a_j$ |
| $\nu^r$ | Dual variable in $\mathcal{L}_{WF}$ for $\sum_{j \in \mathcal{S}} s_j^r = S$ |
| $K$ | Solution constant for the FROP |
| $\kappa_F$ | Price of fairness |

## I. INTRODUCTION

THE advent of alternative energy sources is causing a paradigm change in the operation of the energy grid [1]. It has shifted the generation of electricity away from a few large power plants towards several smaller individual units that are equipped with PV panels and other means to produce electricity from renewable sources. Although at present this energy is typically utilized to meet the individual units' own energy needs, it is envisaged that with greater penetration of PV-equipped homes in future, along with the development of more efficient solar panels, individual homes would be able to deliver energy to the grid [2]. Being positioned closer to other consumer units, these PV-equipped units are better placed to supply energy to the latter during exigent situations [3]. Complete isolation of a microgrid is an extreme example of such a case. Under these circumstances, the microgrid should

This work was supported by the National Science Foundation-CPS under Grants CNS-1136040 and CNS-1544705.

allow bidirectional energy transactions between the units in the form of an auction mechanism that allows the buying and selling of electricity.

*A. Background*

Advances in communications and networking have made automated energy transactions via the grid feasible [1]. Consequently energy trading and grid auction design has been the subject of considerable recent research. Due to the complex nature of the problem, some recent work has focused on the application of nature inspired metaheuristics [4]-[10].Genetic algorithms, swarm intelligence, and hybrid approaches are popular choices for such applications [11].

Linear programming is another popular choice of algorithm. Discrete variables are handled either by means of tree-based search or relaxing and treating them as continuous ones. Unfortunately these approaches entail the assumption of linearity and might not be the ideal choice of many grid auctions [12]-[13].

Approaches for supply side auctions between power generation companies to sell energy at competitive prices have been the subject of much recent research [14]-[22]. Typically these approaches address generation scheduling and unit commitment whose treatment involves discrete design variables. Consequently, mixed integer linear programming has been extensively applied in such studies [15],[19],[22],[23].A two-stage bidding approach between the generation companies and a retailer that procures energy for distribution among consumers has been recently examined [15]. A large-scale day-ahead clearing scheme for the European market has been explored [19]. Mixed integer linear programming to minimize consumer price, while considering generator minimum up/down and ramp up/down times [23], and elsewhere, the presence of shiftable loads (i.e. loads which can be transferred across time slots) [22] have been investigated. Other supply side auctions make approximations in order to use linear programming [16]. One such study pertaining to the Brazilian energy grid, considers piecewise linear utility functions [16]. Linearizing the constraints is used within a game-theoretic equilibrium formulation [18]. A game-theoretic approach for decision making of storage units as seller agents in a smart grid for the maximum amount of energy to sell in the local market so as to maximize a utility that reflects the tradeoff between the revenues from energy trading and the accompanying costs has been studied [20]. A primal-dual approach to obtain optimal power flow is considered within a heterogeneous pricing framework [17]. Unfortunately, approaches where generating companies are involved in the bidding process are inapplicable within our context. Such auctions approach the problem primarily to establish a game-theoretic equilibrium, usually the Stackelberg equilibrium of a sequential game [15], [18], [21], [23].

Demand side auctions with the optimal procurement of energy among multiple buyers is another area of research activity [3], [6], [24]-[27]. A simulation study using a Java based package (JADE) has been carried out [3]. Many of these studies investigate bidding across multiple time frames focusing on optimal operation of shiftable loads [21], [26]. An auction algorithm that incentivizes buyer participation is considered [25]. This study relies on historical data to penalize cheating behavior and ensure truthfulness. A limitation of this study is the underlying assumption of quadratic costs to make the problem formulation strictly convex.

In recent years, double auctions that involve both buyers and sellers of energy, with the latter being PV-equipped units rather than generation companies have begun to be examined [4], [20], [28]-[31]. Unlike auctions between generation companies discussed previously that primarily aim to lower operation costs, the goal of these auction mechanisms is to optimize the distribution of energy within the customers in order to maximize the overall social welfare function (SWF) of the community, *i.e.* the aggregate utilities of all the units of the grid. One study that maximizes SWF, models its consumers as agents that collectively maximize the SWF [32]. This is an unrealistic approach for real world deployment where each agent adjusts is usage patterns only to maximize its individual payoff, *i.e.* the difference between its individual utility from consuming a certain amount of energy and the price it pays to procure the amount. Another study models the agents' utilities as quadratic functions [33].

A few recent studies have implemented the VCG mechanism for energy allocation and trade [24], [34]. This mechanism is used for energy allocation between multiple buying agents [24]. However the approach includes only a single seller; a bottleneck when the grid contains several PV equipped units.

*B. Contribution*

This research proposes a transactive energy *double auction* mechanism in the microgrid containing several domestic units acting as agents and with no restrictions on the number of PV-equipped homes willing to sell energy to other units. In other words, the proposed auction is *multi-agent*, *i.e.* generalized enough to be applicable to systems involving multiple *sellers* as well as *buyers*, which are modeled as sets of distributed agents. In addition, it assumes that there exists a separate agent, the *microgrid controller* (MC), which contains enough computational capabilities to act as an impartial auctioneer.

In contrast to most work on grid auctions which consider uniform pricing across all users in each time interval, this double auction uses *price discrimination*, where individual agents are priced separately. There are only a few papers that use discriminatory pricing [14], [17], [22]; unfortunately none are applicable to the present context.

The proposed auction mechanism is *weakly budget balanced*, so that the total reimbursement provided to the sellers in exchange for energy never exceeds the total revenue obtained from the buyers.

The payoff of each participating agent in the proposed auction is always guaranteed to not be lower than what its payoff would have been from non-participation in the trade. In other words, the auction is *individually rational*.

The proposed mechanism allows separate and arbitrary utility functions for the agents, as long as they are monotonically increasing and concave. Moreover, the PV-equipped sellers have different maximum generation capacities. Although most proposed auction mechanisms make use of this information, in reality it must remain hidden from the MC. The proposed auction is able to attain the desirable outcome without the use of this information. Thus, the proposed auction is *privacy-preserving*. However, it must be noted that the auction may include an optional *redistribution mechanism* that may need access to such information [35], [36]. The redistribution option is included in the proposed mechanism in order to entertain the possibility of further agent coordination beyond the auction. This is when supply is high enough to meet the buyers' demand, leaving room for further bargaining with the sellers, who then opt to impose their own arbitrary *fairness criteria* to redistribute the allocated supply determined by the auction. Such a situation may not arise when there are relatively few sellers, as the outcome of the proposed auction would involve selling their entire surplus energy.

With a high proportion of sellers, the reasons for redistribution are manifold. Agents may have to operate within a legal reimbursement framework [37], resolve conflicts of interest [38]-[40], or other economic incentives [39], [41], [42]. The presence of storage devices either individually, or at a community level is another reason for further redistribution [38], [42]-[45]. Although, for simplicity, this paper considers fair redistribution only for the sellers, the approach can readily be extended to include the buyers.

The proposed auction is SW maximizing for the set of buyers. When the supply is relatively small, the auction also maximizes the sellers' SW. Furthermore, when no extraneous fairness criterion is applied, even with enough supply, the auction is still able to attain the efficient allocation among all agents. However, when there is a need for fair redistribution, the sellers' SW is no longer at its maximum. This aspect of the proposed auction will be referred to as *quasi-efficiency*. The tradeoff between fairness and efficiency is quantified in terms of the *price of fairness*. The fair redistribution mechanism may be incorporated either *in-auction*, or as a second stage, *post-auction* algorithm.

The remainder of this paper is organized as follows. The framework of the auction, and its theoretical analysis are addressed in sections II and III. Fair redistribution is discussed in section IV. Simulation results are presented in section V while section IV concludes this research.

## II. AUCTION FRAMEWORK

The microgrid consists of a set of buyer agents denoted as $\mathcal{D}$ and a set of seller agents denoted as $\mathcal{S}$. At the beginning of the iterative auction, the MC, relays an initial price, which may reflect the actual price under non-isolated operation when the microgrid receives energy from the main grid. In order to ensure weak budget balance, the sellers can sell energy only at prices lower than or equal to $p$, whereas the buyers can procure energy at values higher than or equal to $p$.

Each seller responds to the MC by letting the latter know of the amount of energy $a_j$ available for trade at a per unit price $p_j \leq p$. The energy $a_j$ can never exceed its total energy generation $g_j$. Subsequently the auction proceeds in an iterative manner as shown in Fig. 1.

An iteration of the proposed auction mechanism involves the following exchange of information. The MC computes the volume of energy $s_j \leq a_j$ that it is willing to procure from each seller, and separately $d_i$ that it can deliver to each buyer. The microgrid controller optimization problem (MCOP) used to compute $d_i$ and $s_j$ for this task is addressed later.

The buyer replies to the MC by placing a bid $b_i$ in monetary units that it is willing to pay for $d_i$ units of energy. Note that the condition $pd_i \leq b_i$ for weak budget balance is considered only by the MC. simultaneously, the sellers return $c_j$, the per unit selling cost at which it is willing to supply the amount $s_j$.

As seen in Fig. 1, private information is not provided to the MC. The underlying social welfare optimization problem (SWOP) that ensures efficiency which incorporates both public and private data is discussed first.

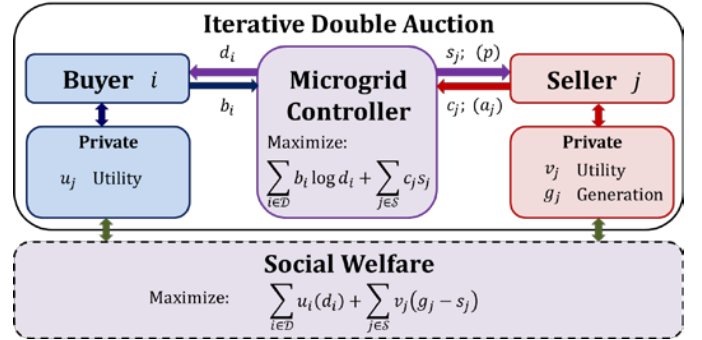

Fig. 1. Schematic showing the flow of information between buying and selling agents and the MC. All parameters except those appearing within parenthesis are updated iteratively.

### A. Social Welfare Optimization Problem

The SWF that is maximized by SWOP consists of the total of all buyers' and sellers' utilities ($u_i$ and $v_j$), summed separately as shown below, where for notational convenience, the arguments $d_i$ and $s_j$ within the function $\Theta(\cdot)$ hereafter refer to the demand and supply allocations for buyers and sellers.

Maximize w.r.t. $d_i, s_j$:

$$\Theta(d_i, s_j) = \sum_{i \in \mathcal{D}} u_i(d_i) + \sum_{j \in \mathcal{S}} v_j(g_j - s_j), \quad (1)$$

subject to:
$$pd_i \leq b_i; \quad i \in \mathcal{D}, \quad (2)$$
$$s_j \leq a_j; \quad j \in \mathcal{S}, \quad (3)$$
$$\sum_{i \in \mathcal{D}} d_i = \sum_{j \in \mathcal{S}} s_j. \quad (4)$$

The first constraint in Eqn. (2) pertains to weakly budget balance for the buyers. The second constraint in Eqn. (3) ensures that the amount of energy that a seller exports to the microgrid never exceeds its declared availability. The last constraint in Eqn. (4) is present to ensure energy balance.

For any given values of the bids $b_i$ and availabilities $a_j$ given by the constraints in Eqns. (2) and (3), the SWOP defines a unique maximum at $d_i^*, s_j^*$. This follows from the fact that the SWOP objective function $\Theta(d_i, s_j)$ is the sum of strictly concave functions, and is also strictly concave with all its constraints being linear. The Lagrangian function corresponding to the SWOP can be written as,

$$\mathcal{L}_\Theta(d_i, s_j, \lambda_i, \alpha_j, \mu) = \Theta(d_i, s_j) + \sum_{i \in \mathcal{D}} \lambda_i(pd_i - b_i)$$
$$+ \sum_{j \in \mathcal{S}} \alpha_j(s_j - a_j) + \mu\left(\sum_{i \in \mathcal{D}} d_i - \sum_{j \in \mathcal{S}} s_j\right), \quad (5)$$

resulting to the following equilibrium conditions
$$pd_i^* \leq b_i, \quad (6)$$
$$\lambda_i^*(pd_i^* - b_i) = 0, \quad (7)$$
$$\alpha_j^*(s_j^* - a_j) = 0, \quad (8)$$
$$u_i'(d_i^*) + \lambda_i^* p + \mu^* = 0, \quad (9)$$
$$-v_j'(g_j - s_j^*) + \alpha_j^* - \mu^* = 0. \quad (10)$$

### B. Microgrid Controller Optimization Problem

In order to achieve the SWOP objective, MCOP is formulated as shown below.

Maximize w.r.t. $d_i, s_j$:
$$\Phi(d_i, s_j) = \sum_{i \in \mathcal{D}} b_i \log d_i - \sum_{j \in \mathcal{S}} c_j s_j, \quad (11)$$

subject to constraints in Eqns. (2), (3), and (4) which are restated below,
$$pd_i \leq b_i; \quad i \in \mathcal{D},$$
$$s_j \leq a_j; \quad j \in \mathcal{S},$$
$$\sum_{i \in \mathcal{D}} d_i = \sum_{j \in \mathcal{S}} s_j.$$

It must be noted that the MCOP formulation does not involve any hidden information from the buyers and sellers. For this reason, the objective that is maximized in MCOP does not involve the agents' utility functions. Likewise, the second MCOP constraint uses $a_j$ instead of $g_j$, the latter being hidden from the MC.

The first term in the MCOP objective function in Eqn. (11), which pertains to the buyers, is adapted from the Kelly mechanism. This mechanism is originally proposed for single sided network auctions [46] which also does not require hidden data. The Kelly mechanism has been studied in the context of communication networks [47]-[49]. The authors have suggested the use of such an auction for use in microgrid energy trade [50]. However, to the best of the authors' knowledge, its use in these auctions has not been examined so far elsewhere.

The second term in the MCOP objective function in Eqn. (11), the summation of the monetary payment $c_j s_j$ given to each seller $j \in \mathcal{S}$, is the total sellers' reimbursement by the MC. In order to accommodate any desired fairness criteria for the sellers, this term has been cast as a linear function. Post-auction redistribution does not change the MCOP objective as long as the reimbursed amount and the total volume of energy transacted do not change during the redistribution stage. For the same reason, in-auction redistribution can be readily incorporated within the proposed mechanism, simply by adding a weighted third term to the objective function.

For simplicity, this research takes into account only seller side fair redistribution. This setup may be viewed as one where the sellers have their own separate arrangement for fair redistribution [51]-[53], while the buyers are conventional consumers of energy. However, the framework can be readily extended to include coordinated buyers. This may be accomplished in a straightforward manner by incorporating another linear term in the MCOP objective, similar to the second but with opposite sign.

The formulation in MCOP offers the flexibility of any redistribution scheme among the sellers using any fairness criterion as long as the total energy volume $S$ supplied by them remains equal to that delivered to the buyers, and the total monetary amount reimbursed to them is fixed. All such solutions satisfying these conditions for $s_j$ must be included in the set of optima of the MCOP. Fig.2 shows a graphical illustration of these considerations. Note that the optimum solution of the MCOP for the buyers, $d_i$, is unique and coincides with that of the SWOP. The sellers' unique optimum solution of the SWOP is also optimal for the MCOP. This solution can be made unique in the SWOP with the inclusion of a third convex term for fairness with a very small weight. Our simulations suggest that, when the auction proceeds without this third term, the auction arrives at the unique SWOP solution.

The Lagrangian of the MCOP is defined as,
$$\mathcal{L}_\Phi(d_i, s_j, \gamma_i, \beta_j, \nu)$$
$$= \Phi(d_i, s_j) + \sum_{i \in \mathcal{D}} \gamma_i(pd_i - b_i) + \sum_{j \in \mathcal{S}} \beta_j(s_j - a_j)$$
$$+ \nu\left(\sum_{i \in \mathcal{D}} d_i - \sum_{j \in \mathcal{S}} s_j\right), \quad (12)$$

resulting to the following equilibrium conditions.
$$pd_i^\dagger \leq b_i, \quad (13)$$
$$\gamma_i^\dagger(pd_i^\dagger - b_i) = 0, \quad (14)$$
$$\beta_j^\dagger(s_j^\dagger - a_j) = 0, \quad (15)$$
$$\frac{b_i}{d_i^\dagger} + \gamma_i^\dagger p + \nu^\dagger = 0, \quad (16)$$
$$-c_j + \beta_j^\dagger - \nu^\dagger = 0. \quad (17)$$

### C. Buyer Bidding

The buyer bids to maximize its own payoff $\pi_i$. This can be formulated as another problem that is carried out locally by the agent.

Maximize w.r.t. $b_i$:
$$\pi_i = u_i(d_i) - b_i. \quad (18)$$

Differentiating w.r.t. $d_i$, yields the following,
$$u_i'(d_i) = \frac{\partial b_i}{\partial d_i}. \quad (19)$$

Upon receiving $d_i$ from the MC, each buyer bids,
$$b_i = u_i'(d_i)d_i. \quad (20)$$

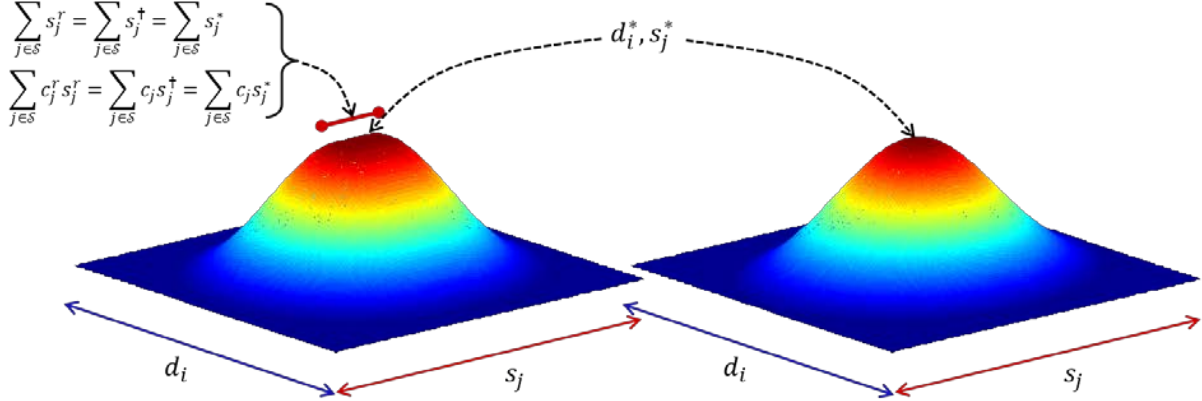

**Fig. 2.** Schematic showing the optima defined by the MCOP (*left*) and the SWOP (*right*). Both optima are unique with respect to the buyers and coincide ($d_i^\dagger, d_i^*$). The SWOP's unique sellers solution ($s_j^*$) is also an optimal solution ($s_j^\dagger$) of the MCOP although the MCOP admits other optima ($s_j^r$) depending on the fairness criterion as long as the constraints shown are satisfied.

*D. Seller Bidding*

At the beginning of the proposed iterative auction, the seller declares its availability $a_j$. The seller communicates the cost $c_j$ at which it is willing to deliver the volume $s_j$ of energy to the microgrid using the following problem formulation.

Maximize w.r.t. $c_j$:
$$\pi_j = v_j(g_j - s_j) + c_j s_j. \quad (21)$$

When the seller does not overbid or underbid, this leads to the following cost updating rule.
$$c_j = v_j'(g_j - s_j). \quad (22)$$

The reason why the seller does not overbid or underbid is as follows. Let us consider the case where $a_j > s_j$. Clearly the seller $j$ would not underbid by declaring a cost $c_j < v_j'(g_j - s_j)$ since the monetary payoff $c_j \Delta s_j$ obtained from this approach would be lower than the loss in utility $v_j(g_j - s_j - \Delta s_j)$. On the other hand, overbidding is not an optimal strategy since it would make $s_j = 0$. This can be seen by inserting the implicit constraint $s_j \geq 0$ to the MCOP problem. In this case, the Lagrangian in Eqn. (12) becomes $\mathcal{L}_\Phi(d_i, s_j, \gamma_i, \beta_j, \nu) + \zeta s_j$ with the KKT conditions $\zeta \geq 0$ and $\zeta s_j = 0$ in addition to those of the MCOP problem given by Eqns. (13) to (16) and with Eqn. (17) replaced with the equality $-c_j + \beta_j + \zeta - \nu = 0$. Since $\beta_j = 0$ when $a_j > s_j$, the equality reduces to $-c_j + \zeta - \nu = 0$. When the seller $j$ does not overbid for a supply $s_j > 0$, it is seen that $\zeta = 0$ and $c_j = -\nu$. However if it overbids, then $c_j > -\nu$ whence $\zeta > 0$ so that $-c_j + \zeta - \nu = 0$, whence the new $s_j$ is forced to be zero, removing the seller from the auction.

When $a_j = s_j$ a similar argument with $\zeta$ replaced with $\beta_j + \zeta$, indicating that the seller will neither overbid nor underbid.

## III. ANALYSIS

This section establishes various desirable features of the proposed transactive energy double auction mechanism.

**Proposition-1** The allocation $d_i^\dagger$ of each buyer $i$ at the maximum of MCOP is equal to the corresponding maximum $d_i^*$ of SWOP, i.e. $d_i^\dagger = d_i^*$.

*Proof*: From the assumption of strict concavity of any buyer's utility $u_i(\cdot)$, the function $u_i'(d_i)d_i$ given by the buyers' bid $b_i$ in Eqn. (20) is strictly increasing. Since the buyer's bid $b_i$ remains unchanged for both allocations $d_i^\dagger$ and $d_i^*$, clearly $u_i'(d_i^\dagger)d_i^\dagger = u_i'(d_i^*)d_i^*$. Hence it follows that $d_i^\dagger = d_i^*$.
∎

**Proposition-2** (*Quasi-efficiency*) The unique SWOP maximum at $d_i^*, s_j^*$ satisfies the KKT conditions of MCOP, so that,
$$(d_i^*, s_j^*) \in \underset{d_i, s_j}{\operatorname{argmax}} \Phi(d_i, s_j).$$

*Proof*: From Proposition-1, $d_i^\dagger = d_i^*$. Consider the case with $s_j^\dagger = s_j^*$. Letting $\gamma_i^\dagger = \lambda_i^*$, $\beta_j^\dagger = \alpha_j^*$, $\nu^\dagger = \mu^*$ and $c_j = -v_j'(g_j - s_j^*)$, Eqn. (13) – Eqn. (17) are satisfied. The statement of Proposition-2 follows immediately. Note that there may exist other values of $s_j^\dagger \neq s_j^*$ satisfying MCOP's KKT conditions so that $(d_i^*, s_j^\dagger) \in \underset{d_i, s_j}{\operatorname{argmax}} \Phi(d_i, s_j)$. This extra degree of freedom offers the option of post-auction sellers' redistribution.
∎

**Proposition-3** (*Weak budget balance*) The proposed auction mechanism is weakly budget balanced.

*Proof*: The net revenue remaining with the MC at the end of the auction is,
$$\pi_{MC} = \sum_{i \in \mathcal{D}} b_i - \sum_{j \in \mathcal{S}} c_j s_j^\dagger. \quad (23)$$

The statement implies that $\pi_{MC} \geq 0$. Hence the following inequality must be established,
$$\sum_{i \in \mathcal{D}} b_i \geq \sum_{j \in \mathcal{S}} c_j s_j^\dagger. \quad (24)$$

The net revenue obtained from the buyers is the bids $b_i$ summed over all buyers, $i \in \mathcal{D}$. Using Eqn. (13) the following inequality holds,

$$\sum_{i \in \mathcal{D}} b_i \geq \sum_{i \in \mathcal{D}} p d_i^\dagger. \tag{25}$$

From the energy balance constraint given by Eqn. (4) at the equilibrium, Eqn. (25) can be written as follow,

$$\sum_{i \in \mathcal{D}} b_i \geq \sum_{i \in \mathcal{D}} p d_i^\dagger = \sum_{j \in \mathcal{S}} p s_j^\dagger. \tag{26}$$

From Eqn. (22) it is seen that $v_j'(g_j - a_j) = c_j$. Since $c_j \leq p$, the inequality in Eqn. (26) can be rewritten as,

$$\sum_{i \in \mathcal{D}} b_i \geq \sum_{j \in \mathcal{S}} v_j'(g_j - a_j) s_j^\dagger. \tag{27}$$

Since $v_j'(g_j - s_j^\dagger) = c_j$ and $s_j^\dagger \leq a_j$, under the assumption that the utilities $v_j(\cdot)$ are concave, $v_j'(g_j - a_j) \geq v_j'(g_j - s_j^\dagger)$. Hence,

$$\sum_{j \in \mathcal{S}} v_j'(g_j - a_j) s_j^\dagger \geq \sum_{j \in \mathcal{S}} v_j'(g_j - s_j^\dagger) s_j^\dagger. \tag{28}$$

From Eqn. (27), Eqn. (28), and Eqn. (22),

$$\sum_{i \in \mathcal{D}} b_i \geq \sum_{j \in \mathcal{S}} v_j'(g_j - s_j^\dagger) s_j^\dagger = \sum_{j \in \mathcal{S}} c_j s_j^\dagger. \tag{29}$$

As $\sum_{j \in \mathcal{S}} c_j s_j^\dagger$ is the reimbursement provided to sellers, the above inequality in Eqn. (29) implies that $\pi_{MC} \geq 0$.
∎

**Proposition-4** (*Individual Rationality*) The proposed auction mechanism is individually rational for all participating agents.
*Proof*: This proposition will be established separately for the buyers and the sellers. Since the bidding strategy of every buyer $i$ is to maximize its payoff $\pi_i = u_i(d_i) - b_i$ where $b_i = u_i'(d_i) d_i$, upon termination of the auction, i.e. at equilibrium, it is evident that,

$$d_i^\dagger = \operatorname{argmax}(u_i(d_i) - u_i'(d_i) d_i). \tag{30}$$

Whence it follows that,

$$u_i(d_i^\dagger) - u_i'(d_i^\dagger) d_i^\dagger \geq u_i(0). \tag{31}$$

Since the utility of the buyer in the absence of any auction would have been $u_i(0)$, that is the right hand side of the inequality in Eqn. (31), it is concluded that the auction is individually rational for the buyers.

The payoff of each seller $j$ after the auction terminates is $\pi_j^\dagger = v_j(g_j - s_j^\dagger) + c_j s_j^\dagger$. Since at $s_j^\dagger$, from Eqn. (22), $c_j = v_j'(g_j - s_j^\dagger)$, the payoff can be expressed as,

$$\pi_j^\dagger = v_j(g_j - s_j^\dagger) + v_j'(g_j - s_j^\dagger) s_j^\dagger. \tag{32}$$

Since the seller's strategy is to maximize its payoff, clearly $\pi_j^\dagger \geq v_j(g_j)$. From the Mean Value Theorem, there exists an $r_j \in (0, s_j^\dagger)$ such that,

$$v_j(g_j) = v_j(g_j - s_j^\dagger) + v_j'(g_j - r_j) s_j^\dagger. \tag{33}$$

From the concavity assumption of the utilities $v_j(\cdot)$, $v_j'(g_j - s_j^\dagger) \geq v_j'(g_j - r_j)$ so that using Eqn. (32) and Eqn. (33),

$$v_j(g_j - s_j^\dagger) + v_j'(g_j - s_j^\dagger) s_j^\dagger \geq v_j(g_j). \tag{34}$$

Since $v_j(g_j)$ represents the payoff of the seller $j$ before the auction, from Eqn. (34), clearly the auction is individually rational for the sellers.
∎

## IV. FAIR REDISTRIBUTION

This section addresses the problem of redistribution of the sellers' allocations using a predetermined fairness criterion. The vast literature of computational mechanism design defines several fairness criteria [54]. However, many such paradigms require sellers' hidden information, i.e. their utility functions, in their formulations, contradicting the underlying assumption of this research that the MC does not have access to the latter.

An in-auction implementation of any redistribution scheme can be readily accomplished by adding a fairness term to the MCOP objective weighted infinitesimally as $\eta F(s_j^r)$, $\eta \ll 1$, so that the auction's properties outlined in the previous section are unaffected. Alternately, it can be implemented post-auction as a second stage of the overall mechanism, which is considered here. The redistribution must be carried out in such a manner that the total amount that the MC provides as reimbursement, $R$, to the sellers must remain unchanged. Hence the redistribution algorithm follows the constraint below.

$$R = \sum_{j \in \mathcal{S}} c_j^r s_j^r = \sum_{j \in \mathcal{S}} c_j s_j^\dagger. \tag{35}$$

In a similar manner, the total energy $S$ supplied by the sellers must remain fixed at that determined prior to redistribution. This is because, from energy balance in Eqn. (4), it must equal the total energy delivered to the buyers. Hence,

$$S = \sum_{j \in \mathcal{S}} s_j^r = \sum_{j \in \mathcal{S}} s_j^\dagger. \tag{36}$$

Lastly, the amount that each seller is allocated after redistribution should not exceed its declared availability, so that,

$$s_j^r \leq a_j; \quad j \in \mathcal{S}. \tag{37}$$

As a representative scheme, we focus on the maximum entropy redistribution [54]. The fair redistribution mechanism's using the maximum entropy criterion is given by,

$$F(s_j^r) = \sum_{j \in \mathcal{S}} \frac{s_j^r}{S} \log \frac{s_j^r}{S}. \tag{38}$$

With Eqns. (36) and (37) as constraints, maximizing $F(s_j^r)$ defines a fair redistribution optimization problem (FROP). The Lagrangian of the FROP is,

$$\mathcal{L}_F(s_j^r, \beta_j^r, \nu^r) = \sum_{j \in \mathcal{S}} \frac{s_j^r}{S} \log \frac{s_j^r}{S} + \sum_{j \in \mathcal{S}} \beta_j^r (s_j^r - a_j) + \nu^r \left( \sum_{j \in \mathcal{S}} s_j^r - S \right), \tag{39}$$

with the following equilibrium conditions,

$$\beta_j^r (s_j^r - a_j) = 0. \tag{40}$$

$$1 + \log \frac{s_j^r}{S} + S\beta_j^r + S\nu^r = 0. \tag{41}$$

This leads to solutions of the form, $s_j^r = K e^{-S\beta_j^r}$, where $K = S e^{-1} e^{-S\nu^r}$. For all sellers with $s_j^r < a_j$, Eqn. (40) shows that $\beta_j^r = 0$, whence $s_j^r = K$. Since $\beta_j^r > 0$ for those sellers that reach their maximum availabilities, the inequality $s_j^r < K$ holds. The redistributed allocations can be stated succinctly as,

$$s_j^r = \min(a_j, K), \qquad (42)$$

with the aggregate energy term constraint leading to the expression,

$$K = S - \sum_{s_j^r = a_j} s_j^r. \qquad (43)$$

This reformulation of the FROP leads to the well-known *water filling algorithm* shown in Fig. 3, and can be readily incorporated within the MC as an algorithm of computational complexity $O(|\mathcal{S}|\log|\mathcal{S}|)$.

The sellers per unit energy costs can be implemented in various ways. For instance, uniform pricing leads to,

$$c_j^r = \frac{1}{S}\sum_{j\in\mathcal{S}} c_j s_j^\dagger. \qquad (44)$$

As mentioned earlier, this redistribution is accompanied by a loss in the overall SW that is expressed in terms of price of fairness, and is given by the following equation

$$\kappa_F = \frac{\Theta(d_i^\dagger, s_j^\dagger) - \Theta(d_i^\dagger, s_j^r)}{\Theta(d_i^\dagger, s_j^\dagger)}. \qquad (45)$$

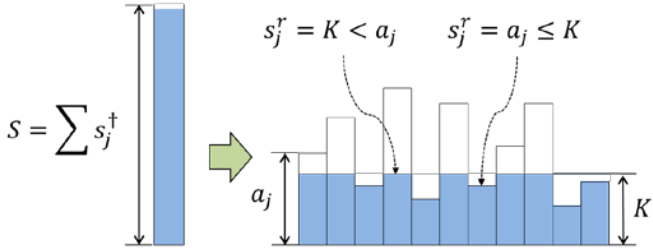

Fig. 3. Illustration of the water filling algorithm. The leftmost column represents the total amount of supply which is redistributed to the columns in the right. In each column, the region shaded in blue (representing water), is the redistributed supply, $s_j^r$.

## V. SIMULATION RESULTS

Several sets of simulations were performed to complement the theory. The auction in every case were initiated with a per unit market price of $p = 0.25$. Utilities of the buyers and sellers were assumed to follow logarithmic saturation curves according to Eqns. (46) and (47),

$$u_i(d_i) = x_i \log(y_i d_i + 1), \qquad (46)$$
$$v_j(g_j - s_j) = x_j \log(y_j(g_j - s_j) + 1). \qquad (47)$$

The quantities $x_i, y_i, x_j$ and $y_j$ were randomly generated for each agent from a uniform distribution centered at unity. The generations, $g_j$, for the sellers were also drawn in at random, uniformly in the interval $[2, 5]$.

In order to show that every individual agent is better off participating in the auction, i.e. the auction is individually rational, extensive simulations were performed to get an average seller and buyers' payoff under two cases of sellers with several cases of buyers as depicted in Fig.4. Notice that as the number of buyers in the auction increases, the average seller's payoff increases while that of the buyer decreases. For a given number of buyers, the payoff of an average seller is higher in the case of 10 sellers than that of 15 sellers and the average payoff of a buyer is lower in the case of 10 sellers than that of 15 sellers.

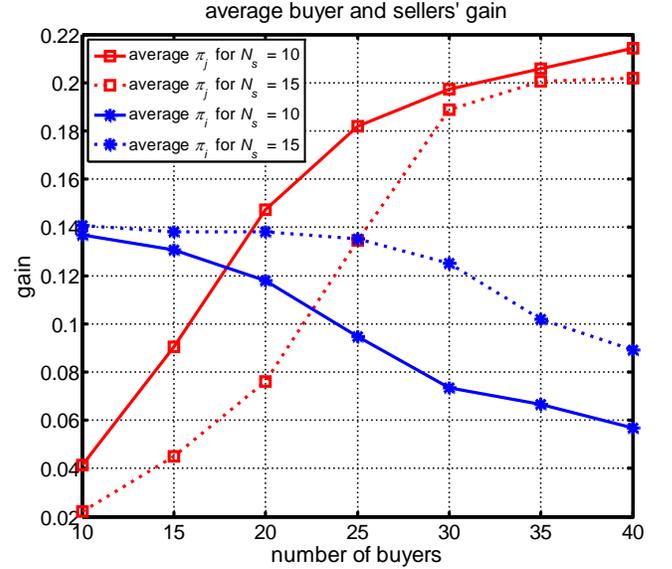

Fig. 4. Average buyer and sellers' payoff participating in the auction for two cases of sellers with several buyers.

To illustrate that the auction allows price differentiation, with $c_i = b_i/d_i$ as the buyers' per unit energy price, Fig. 5 and Fig.6 is presented to show the auction outcome for prices and allocations for two different cases of 5 sellers and 5 buyers (case I) and 5 sellers and 10 buyers (case II) representing two markets with low and high demand. Note that in case I all buyers pay the same minimum per unit price $p = 0.25$ (Fig.5) and receive nonzero allocations (Fig.6) as the number of buyers are lower, whereas they are willing to pay different per unit prices more than $p = 0.25$ in case II as demand is high due to high number of buyers. In case II, buyers who are willing to pay high per unit prices get non-zero allocations. For example, buyers 8, 9, and 10 that are not willing to pay higher per unit prices are allocated zero amounts. Note that in both cases, paying the highest per unit price does not mean getting the highest amount of allocation as every agents' utility curve is randomly generated resulting to different marginal utilities. This means that different agents marginal utilities arrives saturation at different prices after which they are not willing to increase or decrease their bids as it is not profitable.

For the sellers in case I however, except seller 1, all other sellers are allocated lesser supply than their declared availabilities due to low demand in the market, i.e. they end up selling less than their declared availabilities as listed in table I. This is because the buyers' marginal utilities have reached down to saturation at the minimum buying price $p = 0.25$ and they are not allowed to purchase more due to the weakly budget balance constraint $pd_i \leq b_i$. Notice that as there is more supply in the market in case I, the seller(s) with the lowest selling price, i.e. seller 1, gets to sell all its declared availability. Sellers 2 to 4 settle down at almost the same price as that of seller 1 and get to sell most of their declared availabilities whereas seller 5 does not sell any amount due to its high price. For case II however, as the number of buyers is high, the sellers get to sell all their declared availabilities at different per unit prices with seller 5 selling at the highest per unit price.

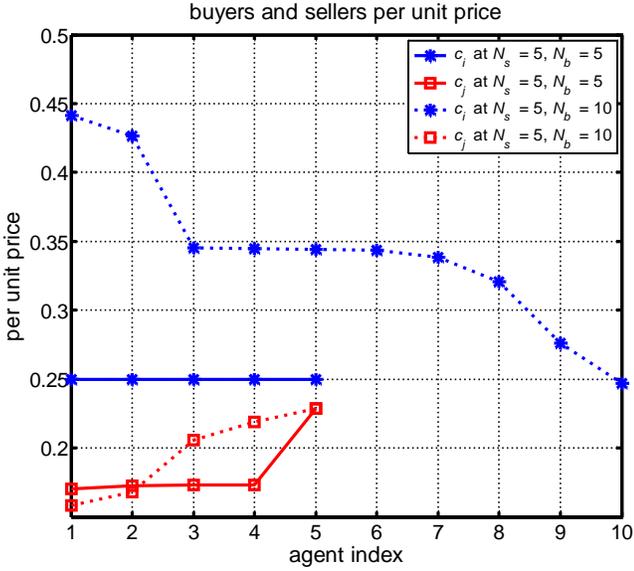

Fig. 5. Buyers and sellers' per unit prices $c_i$ and $c_j$ for $N_s = 5$, $N_b = 5$ (case I) and $N_s = 5$, $N_b = 10$ (case II).

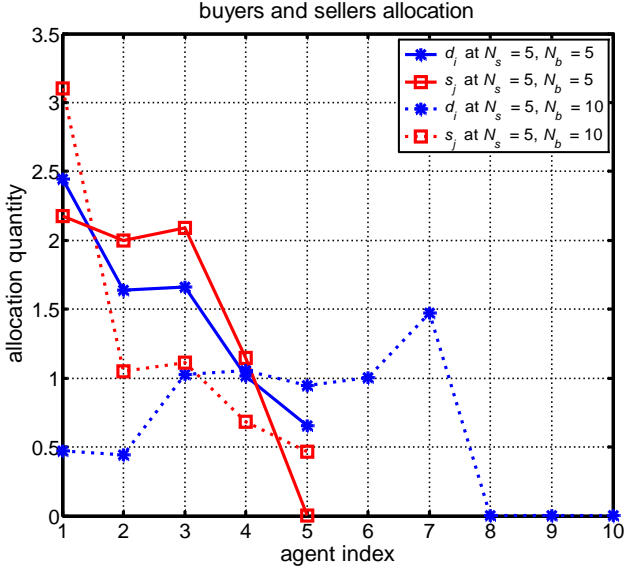

Fig. 6. Buyers' allocation $d_i$ and sellers' allocation $s_j$ for $N_s = 5$, $N_b = 5$ (case I) and $N_s = 5$, $N_b = 10$ (case II).

The MC's revenue $\pi_{MC}$ given by Eqn. (23) in case I is 0.576 whereas it is 1.13 in case II showing the weakly budget balance property of the proposed double auction. This increase, can be readily seen through the change in buyers' per unit prices from case I to case II when they increase from 5 to 10 buyers.

To present the effect of fair redistribution on the sellers' side, additional details of the above two case along with the fair redistribution outputs for the sellers are provided in Table 1. One issue that is solved through a fair redistribution can be observed in case I. Note that sellers 2, 3, and 4 have submitted the same per unit price with their maximum available $a_j$s for sale, however, they have been discriminated during allocation due to multiple optima in the MCOP's objective as illustrated earlier in Fig.2. The water filling algorithm discussed earlier is used for this purpose to fairly redistribute the sellers' allocation, $s_j^r$, with the new equally redistributed per unit price $c_j^r$. This clearly comes with a price, quantified earlier as the price of fairness in Eqn. (45), and is presented later in Fig.7. Notice that in case II, the distribution is already fair in allocations, i.e. $s_j = s_j^r$, as sellers supply at their declared availabilities due to high market demand and redistribution yields the same amounts. However, sellers are price discriminated due to different marginal utilities, which can be redistributed using uniform pricing at sellers consent.

| Case | $j$ | $g_j$ | $a_j$ | $s_j$ | $c_j$ | $s_j^r$ | $c_j^r$ |
|---|---|---|---|---|---|---|---|
| I | 1 | 4.204 | 2.177 | 2.177 | 0.171 | 1.790 | 0.172 |
|   | 2 | 3.205 | 2.022 | 1.997 | 0.173 | 1.790 | 0.172 |
|   | 3 | 3.141 | 2.196 | 2.092 | 0.173 | 1.790 | 0.172 |
|   | 4 | 4.526 | 1.889 | 1.149 | 0.173 | 1.790 | 0.172 |
|   | 5 | 2.155 | 0.254 | 0.000 | 0.229 | 0.254 | 0.172 |
| II | 1 | 4.526 | 3.101 | 3.101 | 0.158 | 3.101 | 0.180 |
|    | 2 | 2.155 | 1.052 | 1.052 | 0.168 | 1.052 | 0.180 |
|    | 3 | 4.204 | 1.112 | 1.112 | 0.206 | 1.112 | 0.180 |
|    | 4 | 3.141 | 0.683 | 0.683 | 0.219 | 0.683 | 0.180 |
|    | 5 | 3.205 | 0.470 | 0.470 | 0.229 | 0.470 | 0.180 |

**TABLE 1** Outcome of the auction pertaining to the sellers, before and after redistribution.

The total sellers' and total buyers' welfare as well as the overall SW under 5 cases when no trade takes place, trade takes place, and when trade takes place and the MC redistributes the allocation for fairness purpose with the associated price of fairness, $\kappa_F$, is illustrated in Fig.7. Note that the SW is higher under trading than the case where no trade takes place, implying the benefit of the auction. Furthermore, the SW decreases after redistribution in the low demand case and is not affected in the high demand cases, where all sellers sell all of their declared availabilities. The price of fairness is only non-zero when some of the sellers do not happen to sell their declared availabilities.

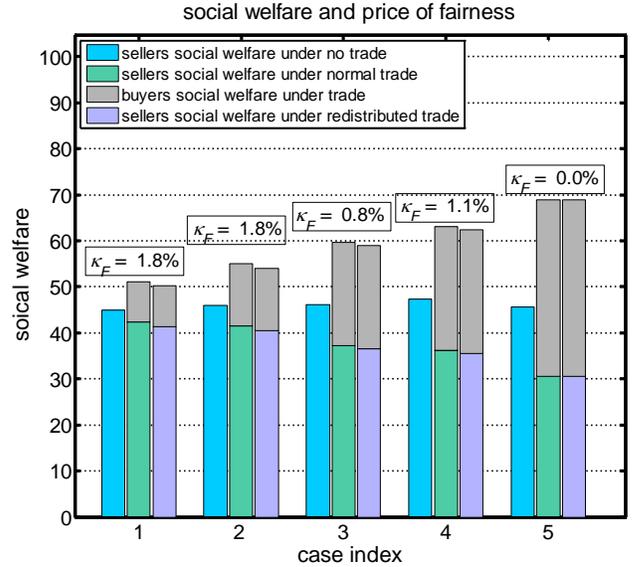

Fig. 7. Social welfare $\Theta(d_i^\dagger, s_j^\dagger)$ under 5 different cases (case 1: $N_s = 50$, $N_b = 20$, case 2: $N_s = 50$, $N_b = 30$, case 3: $N_s = 50$, $N_b = 50$, case 4: $N_s = 50$, $N_b = 60$, case 5: $N_s = 50$, $N_b = 100$) for no trading, trading, and trading with fair redistribution scenarios and the corresponding price of fairness $\kappa_F$ in percent.

Lastly, to show that the auction is efficient, i.e. the MCOP always attains the SW optimum, percent difference of the SW obtained by the MCOP to that of the actual optimum SW has been recorded during each iteration for 4 different cases and has been depicted in Fig.8. As can be seen, the percent difference drops to almost zero within several iterations. Note that the MCOP attains the actual SW optimum given that no in-auction fairness criterion is applied.

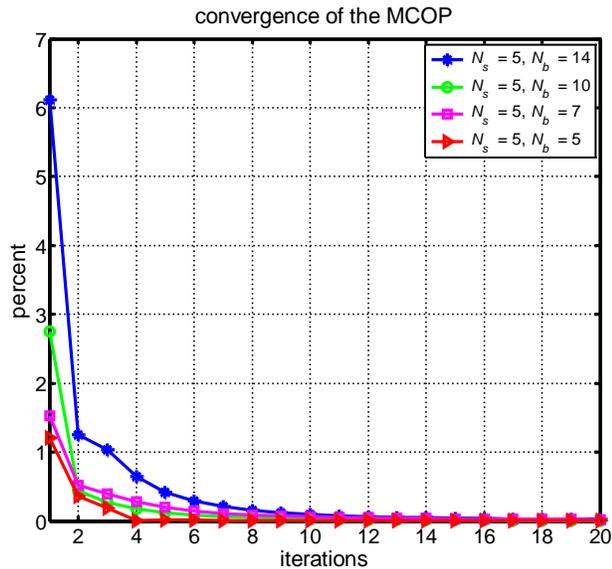

Fig. 8. Difference between SWs attained by SWOP and MCOP as a percentage of the latter for 4 different cases.

## VI. Conclusion

In this research a double sided, weakly budget balanced, individually rational, and efficient transactive energy auction with user hidden information is presented for a microgrid. In the simulations reported earlier, an iteration of the double auction involved multiple steps of the underlying MCOP algorithm in order to ensure that the allocations were close enough to the optima, before the MC allows rebidding. A regularization term weighted by a vanishingly small amount was introduced to MCOP to let it converge to an optimum closest to the initial values. This was done to reduce the communication overhead which varies directly as the number of times the agents rebid. This approach differs from those taken elsewhere [IG+13]. The implicit assumption in this research is that the MC possesses enough processing capabilities to implement an optimization algorithm. However it should be noted that this approach can be implemented in a distribute manner by using methods such as the primal-dual interior point algorithm which may increase the communication steps while reducing the MC's processing requirements [55].

As mentioned earlier, it was observed that without in-auction fairness, the MCOP always converged to the SWOP solution. This was shown in Fig.8 thru percent difference of the SW obtained by the MCOP compared to the actual SW. As theoretical issues pertaining to this observation have not been addressed in this research, the authors do not recommend this approach if no specific fairness criteria are needed.

Given that in double sided auctions, it is impossible to simultaneously achieve perfect efficiency, budget balance, and individual rationality with incentive compatibility[56], [57], in this study the viability of a double sided individually rational, weakly budget balanced, quasi-efficient auction with agents not having to share private user information has been established.

Although not included in our simulation results in the previous section, it is possible to apply water filling fairness criteria in-auction given that the agents' strategies are as given in section II. However, theoretical investigations pertaining to generic in-auction fairness have not been studied.

Future research directions may be directed towards addressing the aforementioned issues.

References

[1] T. Strasser, F. Andrés, J. Kathan, C. Cecati, C. Buccella, P. Siano, P. Leitao, G. Zhabelova, V. Vyatkin, P. Vrba *et al*., "A review of architectures and concepts for intelligence in future electric energy systems," *Industrial Electronics, IEEE Transactions on*, vol. 62, no. 4, pp. 2424–2438, 2015.

[2] E. Sàìz-Marìn, J. Garcìa-Gonzàlez, J. Barquìn, and E. Lobato, "Economic assessment of the participation of wind generation in the secondary regulation market," *Power Systems, IEEE Transactions on*, vol. 27, no. 2, pp. 866–874, 2012.

[3] K. H. Nunna and S. Doolla, "Responsive end-user-based demand side management in multi-microgrid environment," *Industrial Informatics, IEEE Transactions on*, vol. 10, no. 2, pp. 1262–1272, 2014.

[4] B. Ramachandran, S. K. Srivastava, C. S. Edrington, D. Cartes et al., "An intelligent auction scheme for smart grid market using a hybrid immune algorithm," *Industrial Electronics, IEEE Transactions on*, vol. 58, no. 10, pp. 4603–4612, 2011.

[5] J. Nicolaisen, V. Petrov, and L. Tesfatsion, "Market power and efficiency in a computational electricity market with discriminatory double-auction pricing," *Evolutionary Computation, IEEE Transactions on*, vol. 5, no. 5, pp. 504–523, 2001.

[6] R. Herranz, A. Munoz San Roque, J. Villar, and F. A. Campos, "Optimal demand-side bidding strategies in electricity spot markets," *Power Systems, IEEE Transactions on*, vol. 27, no. 3, pp. 1204–1213, 2012.

[7] G. Zhang, G. Zhang, Y. Gao, and J. Lu, "Competitive strategic bidding optimization in electricity markets using bi-level programming and swarm technique," *Industrial Electronics, IEEE Transactions on*, vol. 58, no. 6, pp. 2138–2146, 2011.

[8] Faqiry, M. Nazif, et al. "Game theoretic model of energy trading strategies at equilibrium in microgrids." *North American Power Symposium (NAPS), 2014*. IEEE, 2014.

[9] I. Goroohi Sardou, M. E. Khodayar, K. Khaledian, M. Soleimani-damaneh, and M. T. Ameli, "*Energy and reserve market clearing with microgrid aggregators.*", March 2015.

[10] S. Chakraborty, T. Ito, T. Senjyu, and A. Y. Saber, "Intelligent economic operation of smart-grid facilitating fuzzy advanced quantum evolutionary method," *Sustainable Energy, IEEE Transactions on*, vol. 4, no. 4, pp. 905–916, 2013.

[11] S. Das, "Nelder-mead evolutionary hybrid algorithms." *Encyclopedia of Artificial Intelligence*, vol. 3, pp. 1191–1196, 2009.

[12] S. D. Ramchurn, P. Vytelingum, A. Rogers, and N. R. Jennings, "Putting the'smarts' into the smart grid: a grand challenge for artificial intelligence," *Communications of the ACM*, vol. 55, no. 4, pp. 86–97, 2012.

[13] T. S. Chandrashekar, Y. Narahari, C. H. Rosa, D. M. Kulkarni, J. D. Tew, and P. Dayama, "Auction-based mechanisms for electronic procurement," *IEEE Transactions on Automation Science and Engineering*, vol. 4, no. 3, pp. 297–321, 2007.

[14] A. G. Vlachos and P. N. Biskas, "Simultaneous clearing of energy and reserves in multi-area markets under mixed pricing rules," *Power Systems, IEEE Transactions on*, vol. 26, no. 4, pp. 2460–2471, 2011.

[15] W. Wei, F. Liu, and S. Mei, "Energy pricing and dispatch for smart grid retailers under demand response and market price uncertainty," *Smart Grid, IEEE Transactions on*, vol. 6, no. 3, pp. 1364–1374, 2015.

[16] L. A. Barroso, A. Street, S. Granville, and M. V. Pereira, "Offering strategies and simulation of multi-item iterative auctions of Energy contracts," *Power Systems, IEEE Transactions on*, vol. 26, no. 4, pp.1917–1928, 2011.

[17] N. Alguacil, J. M. Arroyo, and R. Garcia-Bertrand, "Optimization based approach for price multiplicity in network-constrained electricity markets," *Power Systems, IEEE Transactions on*, vol. 28, no. 4, pp. 4264–4273, 2013.

[18] S. J. Kazempour and H. Zareipour, "Equilibria in an oligopolistic market with wind power production," *Power Systems, IEEE Transactions on*, vol. 29, no. 2, pp. 686–697, 2014.

[19] P. N. Biskas, D. Chatzigiannis, A. G. Bakeries et al., "European electricity market integration with mixed market designs part i: Formulation," *Power Systems, IEEE Transactions on*, vol. 29, no. 1, pp. 458–465, 2014.

[20] Y. Wang, W. Saad, Z. Han, H. V. Poor, and T. Basar, "A game theoretic approach to energy trading in the smart grid," *Smart Grid, IEEE Transactions on*, vol. 5, no. 3, pp. 1439–1450, 2014.

[21] Kamyab, Farhad, et al. "Demand response program in smart grid using supply function bidding mechanism." *IEEE Transactions on Smart Grid,* vol.7, no. 3, pp. 1277-1284, 2016.

[22] Kohansal, Mahdi, and Hamed Mohsenian-Rad. "Price-maker economic bidding in two-settlement pool-based markets: The case of time-shiftable loads." *IEEE Transactions on Power Systems,* vol. 31, no. 1, pp. 695-705, 2016.

[23] R. Fernàndez-Blanco, J. M. Arroyo, and N. Alguacil, "Network constrained day-ahead auction for consumer payment minimization," *Power Systems, IEEE Transactions on*, vol. 29, no. 2, pp. 526–536, 2014.

[24] P. Samadi, H. Mohsenian-Rad, R. Schober, and V. W. Wong, "Advanced demand side management for the future smart grid using mechanism design," *Smart Grid, IEEE Transactions on*, vol. 3, no. 3, pp. 1170–1180, 2012.

[25] J. Ma, J. Deng, L. Song, and Z. Han, "Incentive mechanism for demand side management in smart grid using auction," *Smart Grid, IEEE Transactions on*, vol. 5, no. 3, pp. 1379–1388, 2014.

[26] H. Mohsenian-Rad, "Optimal demand bidding for time-shiftable loads," *Power Systems, IEEE Transactions on*, vol. 30, no. 2, pp. 939–951, 2015.

[27] E. Nekouei, T. Alpcan, and D. Chattopadhyay, "Game-theoretic frameworks for demand response in electricity markets," *Smart Grid, IEEE Transactions on*, vol. 6, no. 2, pp. 748–758, 2015.

[28] Faqiry, M. Nazif, and Sanjoy Das. "Double-Sided Energy Auction in Microgrid: Equilibrium under Price Anticipation." *IEEE Access,* vol.4, pp. 3794 – 3805, 2016.

[29] Case, Denise M., et al. "Implementation of a Two-tier Double Auction for On-line Power Purchasing in the Simulation of a Distributed Intelligent Cyber-Physical System." *Advances in Artificial Intelligence*, page 79, 2014.

[30] A. R. Kian, J. B. Cruz, and R. J. Thomas, "Bidding strategies in oligopolistic dynamic electricity double-sided auctions," *IEEE Transactions on Power Systems*, vol. 20, no. 1, pp. 50–58, 2005.

[31] P. Goncalves Da Silva, D. Ilic, and S. Karnouskos, "The impact of smart grid prosumer grouping on forecasting accuracy and its benefits for local electricity market trading," *Smart Grid, IEEE Transactions on*, vol. 5, no. 1, pp. 402–410, 2014.

[32] M. H. Tushar, C. Assi, and M. Maier, "Distributed real-time electricity allocation mechanism for large residential microgrid," *Smart Grid, IEEE Transactions on*, vol. 6, no. 3, pp. 1353–1363, 2015.

[33] P. Samadi, A.-H. Mohsenian-Rad, R. Schober, V.W.Wong, and J. Jatskevich, "Optimal real-time pricing algorithm based on utility maximization for smart grid," in Smart Grid Communications (SmartGridComm), *First IEEE International Conference on*. IEEE, pp. 415–420, 2010.

[34] S. Yang and B. Hajek, "VCG-Kelly mechanisms for allocation of divisible goods: Adapting VCG mechanisms to one-dimensional signals," *in Information Sciences and Systems, 2006 40th Annual Conference on. IEEE*, pp. 1391–1396, 2006.

[35] Buchmann, Erik, et al. "The costs of privacy in local energy markets." *2013 IEEE 15th Conference on Business Informatics*. IEEE, pp. 198-207, 2013.

[36] Kessler, Stephan, Christoph M. Flath, and Klemens Böhm. "Allocative and strategic effects of privacy enhancement in smart grids." *Information Systems,* 53, pp. 170-181, 2015.

[37] Colmenar-Santos, Antonio, et al. "Profitability analysis of grid-connected photovoltaic facilities for household electricity self-sufficiency." *Energy Policy,* 51, pp. 749-764, 2012.

[38] Nykamp, Stefan, et al. "Value of storage in distribution grids—Competition or cooperation of stakeholders?" *IEEE transactions on smart grid* 4.3, pp. 1361-1370, 2013.

[39] Ampatzis, M., Phuong H. Nguyen, and Wil L. Kling. "Introduction of storage integrated PV sytems as an enabling technology for smart energy grids." *IEEE PES ISGT Europe 2013*, 2013.

[40] Nguyen, Phuong H., Wil L. Kling, and Paulo F. Ribeiro. "A game theory strategy to integrate distributed agent-based functions in smart grids." *IEEE Transactions on Smart Grid,* vol.4, no. 1, pp. 568-576, 2013.

[41] Zhao, Jing, Nian Liu, and Jinyong Lei. "Co-benefit and profit sharing model for operation of neighboring industrial PV prosumers." *Smart Grid Technologies-Asia, 2015 IEEE Innovative*, 2015.

[42] Kahrobaee, Salman, et al. "Multiagent study of smart grid customers with neighborhood electricity trading." *Electric Power Systems Research*, 111, pp.123-132, 2014.

[43] Rahbari-Asr, Navid, Yuan Zhang, and Mo-Yuen Chow. "Cooperative distributed scheduling for storage devices in microgrids using dynamic KKT multipliers and consensus networks." *2015 IEEE Power & Energy Society General Meeting*, 2015.

[44] Ampatzis, Michail, Phuong H. Nguyen, and Wil Kling. "Local electricity market design for the coordination of distributed energy resources at district level." *IEEE PES Innovative Smart Grid Technologies, Europe*. IEEE, 2014.

[45] Maity, Indrani, and Shrisha Rao. "Simulation and pricing mechanism analysis of a solar-powered electrical microgrid." *IEEE Systems Journal,* vol.4, no.3, pp. 275-284, 2010.

[46] F. Kelly, "Charging and rate control for elastic traffic," European transactions on Telecommunications, vol. 8, no. 1, pp. 33–37, 1997.

[47] R. J. La and V. Anantharam, "Utility-based rate control in the internet for elastic traffic," *IEEE/ACM Transactions on Networking (TON)*, vol. 10, no. 2, pp. 272–286, 2002.

[48] G. Iosifidis, L. Gao, J. Huang, and L. Tassiulas, "An iterative double auction for mobile data offloading," *in Modeling & Optimization in Mobile, Ad Hoc & Wireless Networks (WiOpt), 2013 11th International Symposium on IEEE,* pp. 154–161, 2013.

[49] W. Tang and R. Jain, "Hierarchical auction mechanisms for network resource allocation," *Selected Areas in Communications, IEEE Journal on*, vol. 30, no. 11, pp. 2117–2125, 2012.

[50] Majumder, B. P., Faqiry, M. N., Das, S., & Pahwa, A. "An efficient iterative double auction for energy trading in microgrids," *Computational Intelligence Applications in Smart Grid (CIASG), 2014 IEEE Symposium on*. IEEE, 2014, pp. 1-7.

[51] Lopez, M. A., et al. "Market-oriented operation in microgrids using multi-agent systems. "*Power Engineering, Energy and Electrical Drives (POWERENG), 2011 International Conference on*. IEEE, 2011, pp. 1-6

[52] Rathnayaka, AJ Dinusha, et al. "Identifying prosumer's energy sharing behaviors for forming optimal prosumer-communities." *Cloud and Service Computing (CSC), 2011 International Conference on*. IEEE, 2011, pp. 199-206.

[53] Kok, Koen. "Multi-agent coordination in the electricity grid, from concept towards market introduction." *Proceedings, 9th International Conference on Autonomous Agents and Multi-agent Systems: Industry track*, 2010, pp. 1681-1688.

[54] Coluccia, Angelo, Alessandro D'Alconzo, and Fabio Ricciato. "On the optimality of max–min fairness in resource allocation." *annals of telecommunications-annales des télécommunications,* 67, no.1-2, pp. 15-26, 2012.

[55] Pakazad, Sina Khoshfetrat, Anders Hansson, and Martin S. Andersen. "Distributed primal-dual interior-point methods for solving loosely coupled problems using message passing." *arXiv preprint arXiv*:1502.06384 (2015).

[56] R. B. Myerson and M. A. Satterthwaite. "Efficient mechanisms for bilateral trading," *Journal of Economic Theory*, 29:265–281, 1983.

[57] D. C. Parkes, J. Kalagnanam, and M. Eso. "Achieving budget-balance with Vickrey-based payment schemes in exchanges," *In Seventeenth International Joint Conference on Artificial Intelligence*, pp. 1161–1168, 2001.